# THERE ARE NO NICE INTERFACES IN 2+1 DIMENSIONAL SOS-MODELS IN RANDOM MEDIA[1]

**Anton Bovier** [2]

*Weierstraß-Institut*
*für Angewandte Analysis und Stochastik*
*Mohrenstrasse 39, D-10117 Berlin, Germany*

**Christof Külske**[3]

*Institut de Recherche Mathématique de Rennes*
*Université de Rennes 1*
*Campus de Beaulieu*
*F-35042 Rennes Cedex, France*

**Abstract:** We prove that in dimension $d \leq 2$ translation covariant Gibbs states describing rigid interfaces in a disordered solid-on-solid (SOS) cannot exist for any value of the temperature, in contrast to the situation in $d \geq 3$. The prove relies on an adaptation of a theorem of Aizenman and Wehr.

---

[1] Work partially supported by the Commission of the European Communities under contract No. CHRX-CT93-0411 and CHBGT-CT93-0244
[2] e-mail: bovier@iaas-berlin.d400.de
[3] e-mail: kuelske@levy.univ-rennes1.fr

# 1. Introduction

In this short note we want to conclude our analysis on the properties of interfaces in random environments by complementing our proof [BK] of the existence of Gibbs measures describing rigid interfaces in the SOS model with random surface tension (at low temperatures and weak disorder) in dimension $d \geq 3$ by showing that on the contrary, in dimension $d \leq 2$, such Gibbs states cannot exist at any temperature as soon as there is any disorder present. In contrast to the technically rather involved existence proof, the proof of the converse statement is simple; in fact it is a fairly straightforward application of a beautiful theorem of Aizenman and Wehr [AW] which they used to prove the uniqueness of the Gibbs state in the two-dimensional random field Ising model. For a extensive discussion of the history of the problem we refer to the introduction of our previous paper [BK].

The model we consider is defined as follows. A surface is described by $\mathbb{Z}$-valued variables $h_x \in \mathbb{Z}$, $x \in \mathbb{Z}^d$. The Hamiltonian is given (formally) by

$$\mathcal{H}(h) = \sum_{<x,y>} |h_x - h_y| + \epsilon \sum_{x,k} \eta_x(k) \mathbb{I}_{h_x = k} \tag{1.1}$$

where $\{\eta_x(k)\}_{x \in \mathbb{Z}^d, k \in Z}$ is a family of independent identically distributed random variables on some abstract probability space $(\Omega, \mathcal{F}, \mathbb{P})$, with non-degenerate distribution $\mathbb{P}$. We assume that $\mathbb{E}[\eta_x(k)] = 0$, $\mathbb{E}[\eta_x(k)^2] = 1$, where $\mathbb{E}$ denotes the expectation w.r.t. the distribution $\mathbb{P}$. As a matter of fact, our result will apply to a far more general class of Hamiltonians but we stick to the specific example for clarity. In [BK] we have proven that under suitable conditions on the temperature and on the distribution $\mathbb{P}$, for $d \geq 3$ infinite volume Gibbs states $\mu_H$ for this model can be constructed as weak limits of finite volume Gibbs measures where the heights on the boundary were set to a fixed and constant value $H$. This reflected the fact that ground states of the Hamiltonian with such boundary conditions tend to be mostly flat interfaces with only rare and localized fluctuations provoked by some large deviations of the random fields. In lower dimensions this is not expected to be the case; rather, on the basis of the Imry-Ma argument [IM], fluctuations are expected to grow without bounds as the volumes increase, resulting in the fact that in the limit as the volume tends to infinity, the probability to observe the interface near the center of the volume at any given height should tend to zero meaning that an infinite volume Gibbs state does not exist. We want to prove a result that reflects this expectation.

To this end we define, following Aizenman and Wehr [AW], the random equivalent of translation invariant Gibbs states, namely *translation covariant* Gibbs states. Let us first note that in the context of random systems, the corresponding random Gibbs measures are most naturally viewed a Gibbs-measure valued random variables on the space $(\Omega, \mathcal{F}, \mathbb{P})$, i.e. a measurable map from $(\Omega, \mathcal{F})$ into the space of Gibbs measures on the measure space of the dynamical variables, in our case $(\mathbb{Z}^{\mathbb{Z}^d}, \mathcal{B})$, where $\mathbb{Z}^{\mathbb{Z}^d}$ is equipped with the product topology of the discrete topology on $\mathbb{Z}$ and $\mathcal{B}$ is the corresponding finitely generated sigma-algebra (a recent exposition on some formal aspects of random Gibbs measures is given in [S]).

**Definition 1:** [AW] *A random Gibbs state $\mu(\eta)$ is called translation covariant iff it satisfies, almost surely,*

(i)
$$\mu\left((\eta_\Lambda(h) + \Delta\eta_\Lambda(h), \eta_{\Lambda^c}(h))_{h \in \mathbb{Z}}\right)(\ \cdot\ )$$
$$= \frac{\mu\left((\eta_\Lambda(h), \eta_{\Lambda^c}(h))_{h \in \mathbb{Z}}\right)(\ \cdot\ \exp(-\beta\epsilon \sum_{x \in \Lambda} \Delta\eta_x(h_x)))}{\mu\left((\eta_\Lambda(h), \eta_{\Lambda^c}(h))_{h \in \mathbb{Z}}\right)(\exp(-\beta\epsilon \sum_{x \in \Lambda} \Delta\eta_x(h_x)))} \quad (1.2)$$

*for any finite volume perturbation $\Delta\eta_\Lambda(h)$ of the random fields, and*

(ii)
$$\mu\left((\eta_{x+y}(h))_{x \in \mathbb{Z}^d, h \in \mathbb{Z}}\right)\left(f(h_x)_{x \in \mathbb{Z}^d}\right)$$
$$= \mu\left((\eta_x(h))_{x \in \mathbb{Z}^d, h \in \mathbb{Z}}\right)\left(f(h_{x-y})_{x \in \mathbb{Z}^d}\right) \quad (1.3)$$

*for all $y \in \mathbb{Z}^d$.*

Let us note that if one translation covariant Gibbs state, say $\mu_0$, exists, than there exists an infinite family of them, $\mu_H$, for all $H \in \mathbb{Z}$, where

$$\mu_H\left((\eta_x(h))_{x \in \mathbb{Z}^d, h \in \mathbb{Z}}\right)\left(f(h_x)_{x \in \mathbb{Z}^d}\right)$$
$$= \mu_0\left((\eta_x(h + H))_{x \in \mathbb{Z}^d, h \in \mathbb{Z}}\right)\left(f(h_x + H)_{x \in \mathbb{Z}^d}\right) \quad (1.4)$$

We will prove the following Theorem:

**Theorem 1:** *Suppose that the distribution $\mathbb{P}$ of $\eta_x(h)$ either*

*(i) has no isolated atoms or*

*(ii) has compact support,*

*then, if $d \leq 2$, $\epsilon \neq 0$, for all $\beta < \infty$, the SOS model defined through (1.1) does not permit translation covariant random Gibbs states.*

**Remark:** Translation covariant Gibbs states are the nice things one expects to get as weak limits with simple boundary conditions which in particular should not be too knowledgeable of the disorder. In particular, property (1.2) can only be violated if $\mu_H$ was constructed as a weak limit with boundary conditions that depended on the random fields in the finite set $\Lambda$. It is quite conceivable that rather artificial Gibbs states violating the conditions (1.2) and (1.3) can be constructed in this model. E.g. it might be possible to choose a sequence of volumes $\Lambda_n \uparrow \mathbb{Z}^d$ and a sequence of random boundary conditions carefully in such a way as to ensure that the corresponding ground states have height $h_0 = 0$ at the origin. It is conceivable that such a sequence of measures could converge, but clearly they are 'physically' irrelevant.

**Remark:** To prove Theorem 1 we will show that the assumption of translation covariant Gibbs states leads to a contradiction. One might hope that a more direct approach based e.g. on the renormalization group method could also work and give more precise information on finite volume quantities. Such an approach, however, appears to be exceedingly difficult. In [K] a result on the absence of stable interfaces based on that idea was proven, but only in a specific mean-field type limit of a hierarchical model. The reader may find it instructive to study that paper, since it hints at the complexities occurring in the problem.

## 2. Proof of the theorem

We will show that the assumption that there exist translation covariant states in $d \leq 2$ leads to a contradiction. Having realized what it is that we want to prove, the adaptation of the arguments of Aizenman and Wehr to our situation is almost trivial. To do so, we define the 'order parameters'

$$M(h, h') \equiv I\!E\left[\mu_0\left(h_x = h\right)\right] - I\!E\left[\mu_0\left(h_x = h'\right)\right] \tag{2.1}$$

The point here is that *if* these quantities vanish, than the we have the following contradiction

$$1 = I\!E \sum_{h \in \mathbb{Z}} \mu_0\left(h_x = h\right) = \sum_{h \in \mathbb{Z}} I\!E \mu_0\left(h_x = h\right) = \sum_{h \in \mathbb{Z}} I\!E \mu_0\left(h_x = h'\right) \tag{2.2}$$

for any $h'$. In fact, if $h^*$ denotes any value for which $I\!E\mu_0\left(h_x = h^*\right) > 0$, to arrive at the same contradiction it is enough to show that there exists an infinite number of values $h$ such that $M(h^*, h) = 0$.

Thus to prove the theorem, we only have to show that this is the case. Let us define, for fixed $h$, $H$, $\beta$, and finite volume $\Lambda$ the generating functions

$$\Gamma_\Lambda(h, H) \equiv \frac{1}{\beta} I\!E \left[\ln \mu_0\left(\exp(\beta\epsilon \sum_{x \in \Lambda} \eta_x(h_x))\right) - \ln \mu_H\left(\exp(\beta\epsilon \sum_{x \in \Lambda} \eta_x(h_x))\right) \bigg| \mathcal{F}_{\Lambda, h}\right] \tag{2.3}$$

where $\mathcal{F}_{\Lambda, h}$ denotes the sigma-algebra that is generated by the random variables $\{\eta_x(h)\}_{x \in \Lambda}$. Define further the random variable

$$\tau_x(h, H) = I\!E\left[\mu_0\left(h_x = h\right) - \mu_H\left(h_x = h\right) \bigg| \mathcal{F}_{\mathbb{Z}^d, h}\right] \tag{2.4}$$

Then we have the following

**Lemma 1:** The functions $\Gamma_\Lambda(h, H)$ and $\tau_x(h, H)$ have the following properties:

(0)
$$\tau_z(h, H)\left((\eta_{x+y}(h))_{x \in \mathbb{Z}^d}\right) = \tau_{z-y}(h, H)\left((\eta_x(h))_{x \in \mathbb{Z}^d}\right) \tag{2.5}$$

(i) For all $x \in \Lambda$
$$\frac{\partial}{\partial \eta_x(h)} \Gamma_\Lambda(h, H) = \epsilon I\!E\left[\tau_x(h, H)\Big|\mathcal{F}_{\Lambda,h}\right] \tag{2.6}$$

(ii)
$$I\!E\left[\tau_x(h, H)\right] = M(h, h - H) \tag{2.7}$$

(iii) For all positive $\beta, \epsilon$,
$$|\tau_x(h, H)| \leq 1 \tag{2.8}$$

and

$$\left|\frac{\partial}{\partial \eta_x(h)} \tau_x(h, H)\right| \leq \frac{\epsilon \beta}{4} \tag{2.9}$$

(iv)
$$I\!E[\Gamma_\Lambda(h, H)] = 0 \tag{2.10}$$

**Proof:** (2.5) follows from (1.3). (2.6) follows from (1.2). (2.7) is a consequence of the 'covariance w.r.t. height shift' expressed by (1.4). The bound (2.8) is obvious. To prove (2.9), just note that

$$\frac{\partial}{\partial \eta_x(h)} \tau_x(h, H) = \epsilon \beta I\!E\left[\mu_0\left(h_x = h\right) - \mu_0\left(h_x = h\right)^2 - \mu_H\left(h_x = h\right) + \mu_H\left(h_x = h\right)^2 \Big|\mathcal{F}_{\mathbb{Z}^d,h}\right] \tag{2.11}$$

(2.10) follows again from (1.4).

◇

Lemma 1 ensures that we are in the situation of [AW] Prop.6.1. which allows us to bound the fluctuations of $\Gamma_\Lambda(h, H)$ from below. In particular we have from Prop.6.1.

$$\liminf_{\Lambda=[-L,L]^d, L\uparrow\infty} I\!E\left[\exp\left(t\Gamma_\Lambda(h, H)/\sqrt{|\Lambda|}\right)\right] \geq \exp\left(\frac{t^2 b^2}{2}\right) \tag{2.12}$$

where (see [AW], Eq.(6.24))
$$b^2 \geq I\!E\left[I\!E\left[\Gamma_\Lambda(h, H)|\mathcal{F}_{0,h}\right]^2\right] \tag{2.13}$$

We distinguish the cases (i) and (ii) in the hypothesis of our theorem. In case (i), Eq. A.3.2 and Proposition A.3.2, case ii. of [AW] immediately gives that

$$b \geq \epsilon \theta_{I\!P}(M(h, H), 1/(\epsilon\beta)) > 0 \tag{2.14}$$

if $M(h, H) \neq 0$.

Under the assumption of case (ii), we show the following

**Lemma 2:** *Let $h^*$ be such that $I\!E\mu_0(h_x = h^*) > 0$. Then there exists $H_0 < \infty$ such that for all $H \geq H_0$*

$$b \geq \epsilon \gamma_{I\!P}(M(h^*, H), 1/(\epsilon\beta)) > 0 \tag{2.15}$$

*if $M(h^*, H) \neq 0$.*

**Proof:** From Proposition A.3.2, case iii. of [AW] (2.15) follows if the function $\eta_0(h^*) \mapsto g(\eta_0(h^*)) \equiv I\!E[\Gamma(h, H)|\mathcal{F}_{0,h^*}](\eta_0(h^*))$ is monotone for all $\eta_0(h^*) \in [A, B]$, where $[A, B]$ is the convex hull of the support of the one field distribution $I\!P$.

To prove the monotonicity, we proceed as follows. From (2.6) we get

$$\begin{aligned}\frac{\partial g}{\partial \eta_0(h^*)} &= \epsilon I\!E\left[\tau_x(h^*, H)\Big|\mathcal{F}_{0,h^*}\right] \\ &= \epsilon I\!E\left[\mu_0(h_x = h^*)\Big|\mathcal{F}_{0,h^*}\right] - \epsilon I\!E\left[\mu_H(h_x = h^*)\Big|\mathcal{F}_{0,h^*}\right]\end{aligned} \tag{2.16}$$

It is easy to see that (1.3) implies the deterministic bounds

$$\delta_H \exp(-2\beta\epsilon(B-A)) \leq I\!E\left[\mu_H(h_x = h^*)\Big|\mathcal{F}_{0,h^*}\right](\eta_0(h^*)) \leq \delta_H \exp(2\beta\epsilon(B-A)) \tag{2.17}$$

for all $\eta_0(h^*) \in [A, B]$ where $\delta_H \equiv I\!E[\mu_H(h_x = h^*)]$. Since

$$\sum_{H \in \mathbb{Z}} \delta_H = \sum_{H \in \mathbb{Z}} I\!E[\mu_0(h_x = h^* - H)] = 1 \tag{2.18}$$

there exists $H_0$ such that $\delta_H \leq \delta_0 \exp(-4\beta\epsilon(B-A))$ for all $H \geq H_0$. This implies by (2.16) the desired monotonicity for all $H \geq H_0$. $\diamondsuit$

To conclude the proof of the theorem, we thus only have to to show that (2.9) with $b^2 > 0$ leads to a contradiction. This relies on the following lemma:

**Lemma 3:**
$$|\Gamma_\Lambda(h, H)| \leq |H||\partial\Lambda| \tag{2.19}$$

*where $|\partial\Lambda| = |\{(x, y)|x \in \Lambda, y \in \Lambda^c, |x - y|_2 = 1\}|$.*

In fact, (2.19) implies

$$I\!E\left[\exp\left(t\Gamma_\Lambda(h,H)/\sqrt{|\Lambda|}\right)\right] \leq \exp\left(|t||H|\frac{|\partial\Lambda|}{\sqrt{|\Lambda|}}\right) \tag{2.20}$$

which is the desired contradiction if $\Lambda$ is chosen as e.g. a $d$-dimensional cube and $d \leq 2$. This concludes the proof of the theorem, if we assume Lemma 3. $\Diamond$

To conclude we prove Lemma 3.

**Proof:** (of Lemma 3) We focus on one summand in (2.2) and write

$$\begin{aligned}
&I\!E\left[\ln\mu_H\left((\eta_\Lambda(h),\eta_{\Lambda^c}(h))_{h\in\mathbb{Z}}\right)\left(\exp(\beta\epsilon\sum_{x\in\Lambda}\eta_x(h_x))\right)\bigg|\mathcal{F}_{\Lambda,h}\right]\\
&= -I\!E\left[\ln\mu_H\left((0_\Lambda,\eta_{\Lambda^c}(h))_{h\in\mathbb{Z}}\right)\left(\exp(-\beta\epsilon\sum_{x\in\Lambda}\eta_x(h_x))\right)\bigg|\mathcal{F}_{\Lambda,h}\right]\\
&= -I\!E\left[\ln\mu_H\left((0_\Lambda,\eta_{\Lambda^c}(h-H))_{h\in\mathbb{Z}}\right)\left(\exp(-\beta\epsilon\sum_{x\in\Lambda}\eta_x(h_x))\right)\bigg|\mathcal{F}_{\Lambda,h}\right]\\
&= -I\!E\left[\ln\mu_0\left((0_\Lambda,\eta_{\Lambda^c}(h))_{h\in\mathbb{Z}}\right)\left(\exp(-\beta\epsilon\sum_{x\in\Lambda}\eta_x(h_x+H))\right)\bigg|\mathcal{F}_{\Lambda,h}\right]
\end{aligned} \tag{2.21}$$

where the first equality is due to the transformation law (1.2) w.r.t. local perturbations, the second to the stationarity of the distribution of the random fields under the shift $h_x \mapsto h_x + H$ for $x \in \Lambda^c$, the third to (1.4).

Let us now employ the DLR-equations (see [Ge]) to write

$$\mu_0\left((0_\Lambda,\eta_{\Lambda^c}(h))_{h\in\mathbb{Z}}\right)\left(\exp(-\beta\epsilon\sum_{x\in\Lambda}\eta_x(h_x+H))\right) = \int \mu_0\left((0_\Lambda,\eta_{\Lambda^c}(h))_{h\in\mathbb{Z}}\right)(d\bar{h}_{\Lambda^c}) \times$$

$$\times \frac{\sum_{h_\Lambda\in\mathbb{Z}^\Lambda}\exp\left(-\beta\sum_{\substack{<x,y>\\x,y\in\Lambda}}|h_x-h_y|-\beta\sum_{\substack{<x,y>\\x\in\Lambda,y\in\Lambda^c}}|h_x-\bar{h}_y|-\beta\epsilon\sum_{x\in\Lambda}\eta_x(h_x+H)\right)}{\sum_{h_\Lambda\in\mathbb{Z}^\Lambda}\exp\left(-\beta\sum_{\substack{<x,y>\\x,y\in\Lambda}}|h_x-h_y|-\beta\sum_{\substack{<x,y>\\x\in\Lambda,y\in\Lambda^c}}|h_x-\bar{h}_y|\right)} \tag{2.22}$$

Note that only the numerator is $H$-dependent. Therefor we introduce $h'_x = h_x + H$ for $x \in \Lambda$ estimate the boundary term in the 'surface-energy' in the exponential in the numerator uniformly by

$$\sum_{\substack{<x,y>\\x\in\Lambda,y\in\Lambda^c}}|h_x-\bar{h}_y| = \sum_{\substack{<x,y>\\x\in\Lambda,y\in\Lambda^c}}|h'_x-\bar{h}_y-H|$$

$$\leq \sum_{\substack{<x,y>\\x\in\Lambda,y\in\Lambda^c}}|h'_x-\bar{h}_y| + |H||\partial\Lambda| \tag{2.23}$$

and

$$\sum_{\substack{<x,y>\\x\in\Lambda,y\in\Lambda^c}}|h_x-\bar{h}_y| \geq \sum_{\substack{<x,y>\\x\in\Lambda,y\in\Lambda^c}}|h'_x-\bar{h}_y| - |H||\partial\Lambda| \tag{2.24}$$

From this we have

$$\sum_{h_\Lambda \in \mathbb{Z}^\Lambda} \exp\left(-\beta \sum_{\substack{<x,y> \\ x,y \in \Lambda}} |h_x - h_y| - \beta \sum_{\substack{<x,y> \\ x \in \Lambda, y \in \Lambda^c}} |h_x - \bar{h}_y| - \beta\epsilon \sum_{x \in \Lambda} \eta_x(h_x + H))\right)$$

$$\leq \exp(\beta |H||\partial \Lambda|) \sum_{h'_\Lambda \in \mathbb{Z}^\Lambda} \exp\left(-\beta \sum_{\substack{<x,y> \\ x,y \in \Lambda}} |h'_x - h'_y| - \beta \sum_{\substack{<x,y> \\ x \in \Lambda, y \in \Lambda^c}} |h'_x - \bar{h}_y| - \beta\epsilon \sum_{x \in \Lambda} \eta_x(h'_x))\right) \quad (2.25)$$

and a similar lower bound. Substituting these bounds in (2.12) and comparing the $H = 0$-term gives (2.10) directly. $\diamond$

To summarize the gist of the proof, Lemma 3 roughly the fact that when we deform a interface aver a local region $\Lambda$ by shifting it up by a distance $H$, then this 'costs' no more than to build a boundary wall, i.e $H|\partial L|$. On the other hand, the Aizenman-Wehr theorem says that there are always regions around where such a price is compensated by a corresponding gain in random energy. In that sense, the proof really builds along the Imry-Ma argument. On the other hand, we see that to make this argument rigorous, one has to proceed quite carefully in order to avoid possible pathologies that could be produced by very "exotic' constructions of Gibbs states. This somewhat restricts the generality of our statement (namely that we only exclude translation covariant Gibbs states rather then 'any' Gibbs states) but such a restriction does not appear physically unreasonable.

# References

[AW] M. Aizenman, and J. Wehr, Rounding effects on quenched randomness on first-order phase transitions, Commun. Math. Phys. **130**: 489 (1990).

[BK] A. Bovier and Ch. Külske, A rigorous renormalization group method for interfaces in random media, Rev. Math. Phys. **6**: 413-496 (1994).

[Ge] H.-O. Georgii, Gibbs measures and phase transitions, Walter de Gruyter (de Gruyter Studies in Mathematics, Vol. 19), Berlin-New York, 1988.

[IM] Y. Imry and S. Ma, Random-field instability of the ordered state of continuous symmetry, Phys. Rev. Lett. **35**:1399-1401 (1975).

[K] Ch. Külske, Interfaces in stochastic media, to appear in Proceedings of the conference 'Advanced topics in applied mathematics and theoretical physics: Complex Systems', Marseille 1994.

[S] T. Seppäläinen, Entropy, limit theorems, and variational principles for disordered lattice systems, to appear in Commun. Math. Phys. (1995).